\documentclass[twocolumn,showpacs,amsmath,amssymb,epsfig]{revtex4}

\usepackage{graphicx}
\usepackage{bm}

\usepackage{epsfig}
\makeatletter
\newcommand\figcaption{\def\@captype{figure}\caption}
\newcommand\tabcaption{\def\@captype{table}\caption}
\makeatother

\begin{document}
\title{Critical Scaling of Extended Power Law I-V Isotherms near Vortex Glass Transition}
 \draft
\author{Z.H. Ning, X. Hu, K.X. Chen, L. Yin, G. Lu, X.L. Xu, J.D. Guo, F.R. Wang , C.Y. Li and D.L. Yin}
\address{Department of Physics, Peking University, Beijing 100871, China}
\date{\today}

\begin{abstract}
In view of the question about the vortex glass theory of the
freezing of disordered vortex matter raised by recent experimental
observations we reinvestigate the critical scaling of high $T_c$
superconductors. We find that dc current-voltage characteristic of
mixed state superconductors has the general form of extended power
law which is based on the Ginzburg-Landau (GL) functional in the
similar way as the vortex glass theory. Isotherms simulated from
this nonlinear equation fit the experimental I-V data of Strachan
et al.[Phys.Rev.Lett. 87, 067007 (2001)]. The puzzling question of
the derivative plot for the I-V curves and the controversy
surrounding the values of critical exponents are also discussed.

\end{abstract}
\pacs{74.25.Sv,74.25.Fy, 74.25.Ha,74.20.De}

\maketitle

The hypothesis of a vortex glass in disordered high temperature
superconductors\cite{R1} has spurred much research and many
discussions during more than one decade and continued to be a very
controversial issue\cite{R2,R3}. Though the striking dc
current-voltage(I-V) data collapse shown by Koch et al.\cite{R4},
and found by other groups on other samples, strongly supports the
consensus that a glass transition occurs in high-temperature
superconductors(HTS)\cite{R5}, this widely accepted evidence has
met serious doubt recently. Strachan et al. show wide range
accurate isothermal I-V measurements over 5 or 6 decades on a high
quality 2200 $\text{\AA}$ thick $YBa_{2}Cu_{3}O_{7-\delta}$ film
with $T_{c}\approx 91.5K$ and transition width about 0.5K\cite{R3}
and find although the I-V isotherms measured in a magnetic field
can be collapsed onto scalling functions proposed by Fisher et
al.\cite{R1} as is widely reported in the literature, these
excellent data collapse can also be achieved for a wide range of
exponents and glass temperatures $T_{g}$ as demonstrated in their
Fig.2(a),2(b) and 2(c)\cite{R3}. Since the critical temperature
$T_g$ cannot be determined uniquely, they argued that necessary
evidence for a vortex-glass transition to zero resistance is still
not seen in their I-V experimental data, as well as in all the
ones they know of in the literature which scale. Therefore, the
correctness of the vortex glass picture has to be reinvestigated.
In this work, we try to settle this remarkable controversy by
introducing a general extended power law form of the
current-voltage isotherms. \\\indent In analogue with the vortex
glass theory \cite{R1}, our extended power law equation is also
based on the GL free energy functional of a system of $N$ flux
lines with a field $H$ along the $z$ direction in a sample length
$L$ which can be described with the free energy represented by the
trajectories $\left\{ \overrightarrow{r_j}%
\left( z\right) \right\} $ of these flux lines and the pinning
potential $V_P\left( \overrightarrow{r}%
\right) $ arising from inhomogeneities and defects in sample
\cite{R1,R5,R6,R7} as\vspace{-12pt}
\begin{eqnarray}\label{1}
F & = & \frac 12\varepsilon _l\sum_{j=1}^N\int_0^L\left| \frac{d\stackrel{%
\rightharpoonup }{r_j}\left( z\right) }{dz}\right| ^2dz+\frac
12\sum_{i\neq j}\int_0^LV\left( r_{ij}\right) dz\nonumber\\
& + & \sum_{j=1}^N\int_0^LV_P\left[ \stackrel{%
\rightharpoonup }{r_j}\left( z\right) \right]dz
\end{eqnarray}
Here $V\left( r_{ij}\right) =V\left( \left| \overrightarrow{r_i}-%
\overrightarrow{r_j}\right| \right) =2\varepsilon _0K_0\left(
r_{ij}/\lambda _{ab}\right) $ is the interaction potential between
lines with in-plane London penetration depth $\lambda _{ab}$ and
$K_0\left( x\right) $ is the modified Bessel function $K_0\left(
x\right) \approx \left( \pi /2x\right) ^{1/2}e^{-x}$. $\varepsilon
_l$ is the linear tension of flux line and $\varepsilon _0\approx
\left( \Phi _0/4\pi \lambda _{ab}\right) ^2$ is the energy scale
for the interaction.

Thermally activated flux motion can be considered as the sequence
of thermally activated jumps of the vortex segments or vortex
bundles between the metastable states generated by disorder. Every
elementary jump is viewed as the nuclearation of a vortex loop,
and the mean velocity of the vortex system is determined by the
nuclearation rate \cite{R1,R7}
\begin{equation}\label{2}
v\varpropto \exp \left(- \delta F/kT\right)
\end{equation}
Here $\delta F$ is the free energy for the formation of the
critical size loop or nucleus which can be found by means of the
standard variational
procedure from the free energy functional due to the in-plane displacement $%
\overrightarrow{u}\left( z\right) $ of the moving vortex during
loop formation
\begin{equation}\label{3}
F_{loop}\left[ \overrightarrow{u}\right] =\int dz\left[ \frac 12%
\varepsilon _l\left| \frac{d\overrightarrow{u}\left( z\right) }{%
dz}\right| ^2+V_P(\overrightarrow{u}(z))-\overrightarrow{f}_s \cdot \overrightarrow{u}%
\right]
\end{equation}
with
\begin{equation}\label{4}
\overrightarrow{f}_s=\overrightarrow{f}_L+\overrightarrow{f}_\eta
=\frac{J_P\Phi _0}c\times \overrightarrow{e_z}
\begin{array}{lll}
&  &
\end{array}
\text{and }
\begin{array}{lll}
&  &
\end{array}
J_P=J-\frac E{\rho _f}
\end{equation}
where $f_L=\overrightarrow{J}\times \overrightarrow{%
e_z}/c$ is the Lorentz force due to applied current $J$ and
$f_\eta $ is the
viscous drag force on vortex, $f_\eta =-\eta v_{vortex}$, with $v_{vortex}=d%
\overrightarrow{u}/dt$ and viscous drag coefficient $\eta \approx
\left( \Phi _0B_{c2}\right) /\left( \rho _nc^2\right) $ as
estimated by Bardeen and Stephen \cite{R8}.

There are two essential differences between our equations (1) and
(3) and the corresponding ones used in previous vortex glass
theories:

(i)We considered both the Lorentz force $\overrightarrow{f}_L$ and
the viscous drag force $\overrightarrow{f}_\eta$ on the vortex
loop during its formation while in present vortex glass theories
 only bare Lorentz force $\overrightarrow{f}_L$ has been taken into
 account\cite{R1,R5,R7}.

(ii)The effect of the real sample size L on the current-voltage
response is explicitly estimated in our case. The influence of the
finite size effects on the I-V isotherms is recently revealed even
in thickest films at zero magnetic field\cite{R9}.

These two factors result in a new form for the I-V isotherms of
superconducting vortex matter.

In stead of the barrier energy found by the vortex glass
model\cite{R1} as
\begin{equation}\label{5}
\delta F=U(J)\approx U_c(\frac{J_c}{J})^{\mu}
\end{equation}
now we have the barrier energy
\begin{equation}\label{6}
\delta F=U( J_p) \approx U_{c}(\frac{J_{c}}{J_{p}})^{\mu}
\end{equation}
which implies a current-voltage characteristic of the form
\begin{equation}\label{7}
E(J)=\rho_{f}Jexp[-\frac{U_{c}}{kT}(\frac{J_{c}}{J_{p}})^{\mu}]
\end{equation}
where $U_{c}$ is a temperature- and field-dependent characteristic
pinning energy related to the stiffness coefficient and $J_{c}$ is
a characteristic current density related to $U_{c}$ ,$\mu$ is an
numerical exponent.

Considering the real size effect Eq.(7) leads to a general
normalized form of the current-voltage characteristic \cite{R10}
in the form
\begin{equation}
y=x\exp \left[ -\gamma \left( 1+y-x\right) ^p\right]
\end{equation}
where
\begin{equation}\label{8}
\gamma =\frac{U_c}{kT}\left( \frac{J_c}{J_L}\right) \text{, }
\begin{array}{lll}
&  &
\end{array}
x=\frac J{J_L}\text{, }
\begin{array}{lll}
&  &
\end{array}
y=\frac{E\left( J\right) }{\rho _fJ_L}\text{, }
\begin{array}{lll}
&  &
\end{array}
p=\mu\\\nonumber
\end{equation}
with $J_{L}$ the transport current density corresponding to the
case where the critical size of loop formation is equal to the
real sample size L. The $\rho_{f}$ in Eq.(7) is the flux flow
resistance of a pinning free mixed state as derived by Bardeen and
Stephen. Its relation with the normal resistance is of the form
\begin{equation}\label{9}
\frac{\rho _f}{\rho _n}=\frac{2\pi a^2B}{\Phi _0}=\left( \frac
a\xi \right) ^2\frac B{B_{c2}}\approx \frac B{B_{c2}}
\end{equation}
In an earlier work \cite{R11}, equation(8) has also been shown in
connection with the Anderson-Kim model. Compared with the widely
used power law nonlinear response $E\propto J^{n}$ (see Ref.5),
equation(8) can also be nearly equivalently expressed in an
extended power law form
\begin{equation}\label{10}
\frac{y}{x}= e^{-\gamma}(x-y+1)^{n-1}
\end{equation}
with n a numerical factor depending on the parameters $\gamma$ and
p in Eq.(8), which can be determined from solving the equation
\begin{eqnarray}\label{14}
& &\frac{y_i}{x_i}=e^{-\gamma}(x_i-y_i+1)^{n-1}
\end{eqnarray}
where the subscript $i$ denote the corresponding values at the
inflection point of $lny\sim lnx$ curve for Eq.(8) with the
maximum derivative $S_{max}$ so
\begin{equation*}
\frac{\partial^{2}lny}{\partial^{2}lnx}|_{x=x_{i}}=0
\begin{array}{lll}
& \text{and} &
\end{array}
S_{max}=\frac{\partial lny}{\partial lnx}|_{x=x_{i}}
\end{equation*}

In Fig.1 we show the numerical solutions of Eq.(8) and Eq.(10) for
comparison.

\begin{figure}
\centering
\includegraphics[width=1.0\linewidth]{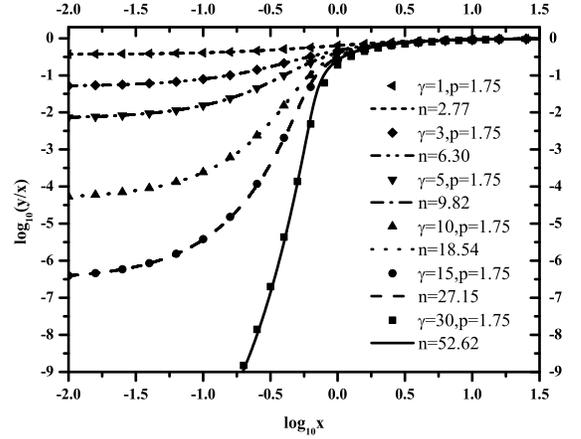}
\caption{Comparison of solutions of Eq.(8)(solid symbols) and
Eq.(10)(curves).}
\end{figure}
The complex experimental current-voltage data are usually
described phenomenologically with a form\cite{R5}
\begin{equation}\label{12}
E(J)=J\rho_{f}(T,B,J)=J\rho_{f}e^{-U_{eff}(T,B,J)/kT}
\end{equation}
Equation(8) implies that effective barrier can be explicitly
expressed as
\begin{equation}\label{13}
U_{eff}(T,B,J)=U_{c}(B,T)F[J/J_{c}(B,T)]
\end{equation}
with$J_{c}(B,T)=J_{L}(B,T)$.

Incorporating it into the commonly observed scaling behavior of
magnetic hysteresis $M(H)$ in superconductors, it can be shown
that $U_{c}(B,T)$ and $J_{c}(B,T)$ in Eq.(13) must take the
following forms\cite{R5}
\begin{eqnarray}\label{14}
U_{c}(B,T)=\Psi(T)B^{n}\propto[T^{*}(B)-T]^{\delta}B^{n}\nonumber\\
J_{c}(B,T)=\lambda(T)B^{m}\propto[T^{*}(B)-T]^{\alpha}B^{m}
\end{eqnarray}
with $T^{*}(B)$ being the irreversibility temperature where tilt
modules and stiffness of vortex matter vanish. Considering
$\rho_{f}\propto T$, one finds from Eq.(8) the following
temperature dependent relations for $\gamma$ and $I-V$ isotherms
of optimally doped YBCO samples in a given magnetic field as
\begin{eqnarray}\label{15}
& &\gamma(T)=\gamma_{0}(T^{*}-T)^{\delta}/kT\nonumber\\
& &I\propto xJ_{L}(T)\propto x(T^{*}-T)^{\alpha}\nonumber\\
& &V\propto yJ_{L}(T)\rho_{f}\propto y(T^{*}-T)^{\alpha}T
\end{eqnarray}
Combination of equations (8) and (15) enables us to simulate the
experimentally measured $I-V$ isotherms of Strachan et
al.\cite{R3} as shown in Fig.2, where the parameters used in (8)
and (15) are $\alpha=3.0, \delta=0.5, p=1.75,
\gamma_{0}=3.21\times10^{21}$ and the irreversibility temperature
$T^{*}(B=4T)$ is assumed $88K$ similar to the Fig.1 in Ref.[3].
\begin{figure}[btp]

\centering

\includegraphics[width=0.7\linewidth,angle=270]{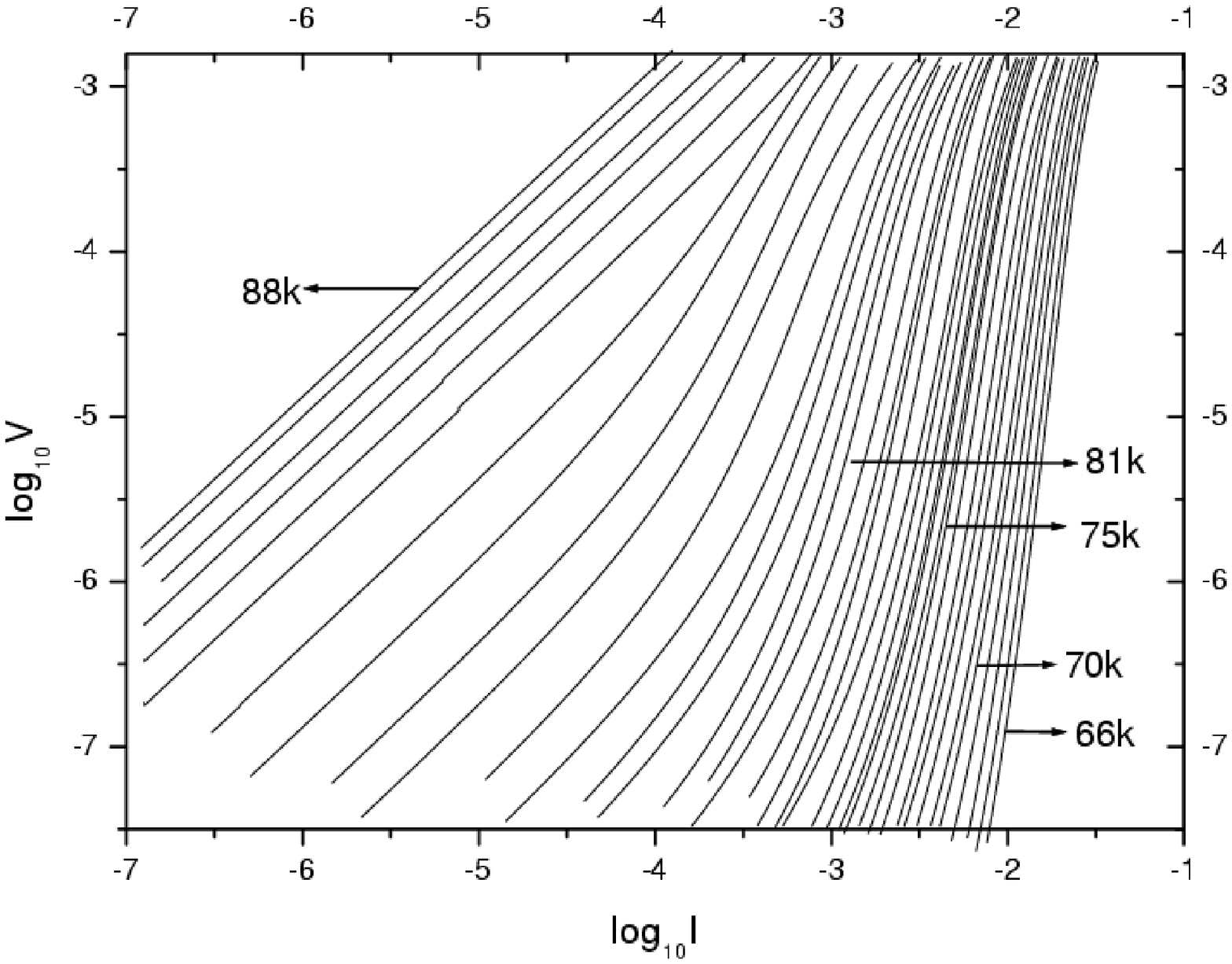}

(a)

\includegraphics[width=0.85\linewidth]{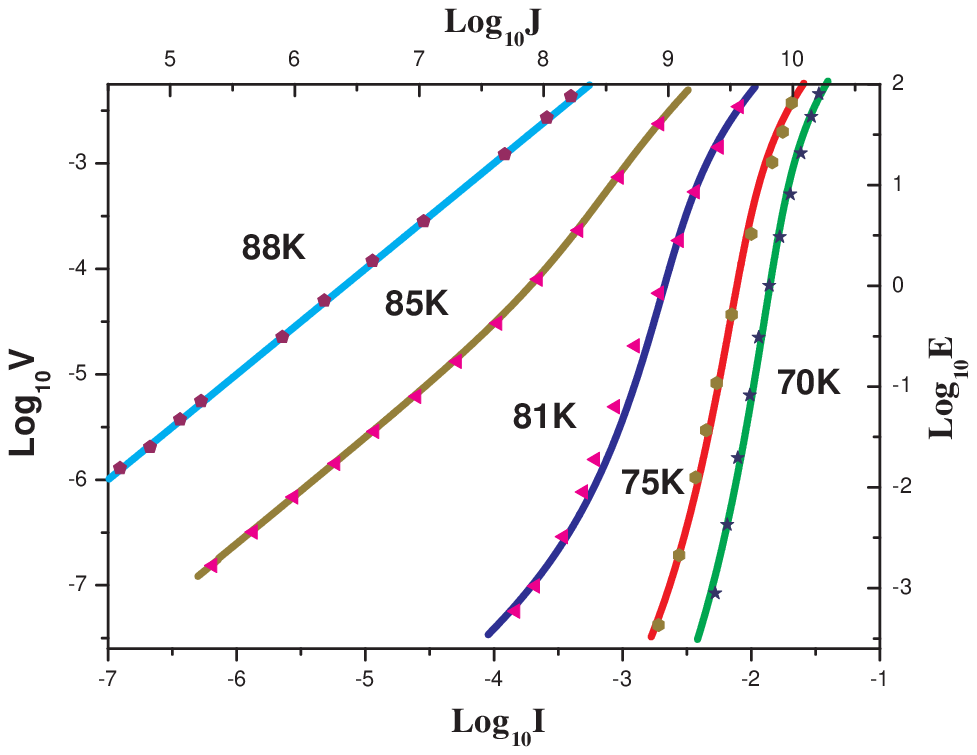}

(b)

\caption{Comparison of simulated I-V isotherms from Eqs.(8) and
(15) with the experimental data of Ref.3[Phys.Rev.Lett.87,
067007(2001)]. \\(a)I-V isotherms simulated from the extended
power law equations (8) and (15) with assumed
$T^{\ast}(B=4T)=88K$, and $\alpha=3.0, \delta=1.5, p=1.75$ and
$\gamma_0=3.21\times10^{21}$. \\(b)Comparison of our simulated I-V
isotherms(lines) with the experimental data of Ref.3(solid
symbols).}

\end{figure}
On the basis of nearly 100 simulated $I-V$ isotherms from $68K$ to
$88K$ with temperature intervals of $0.2K$, we obtain three
different scaling data collapse analyses with assumed critical
temperature $T_{g}$ of $81K$,$75K$ and $70K$
\begin{figure}

\centering

\includegraphics[width=0.585\linewidth,angle=270]{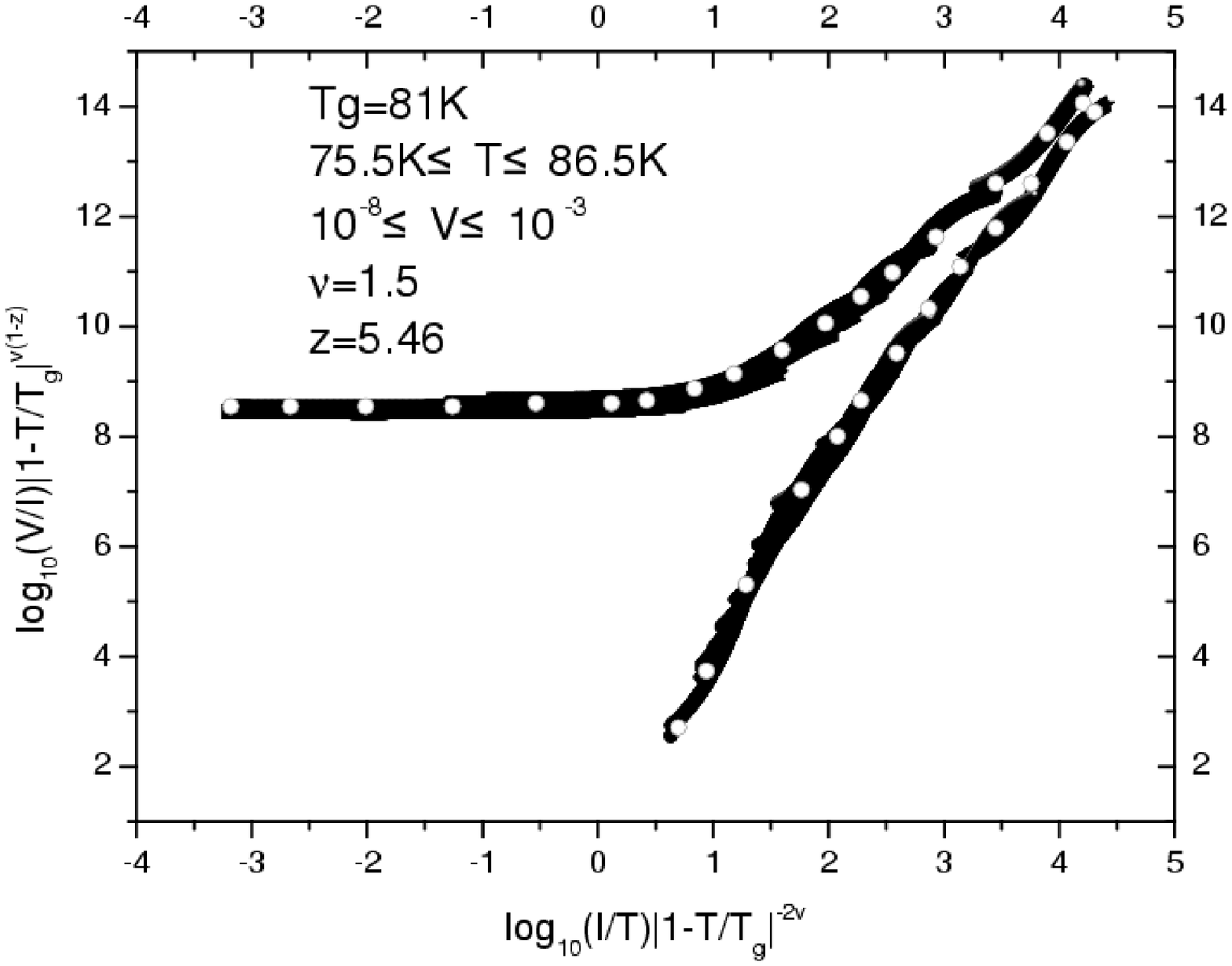}

\noindent (a)

\includegraphics[width=0.585\linewidth,angle=270]{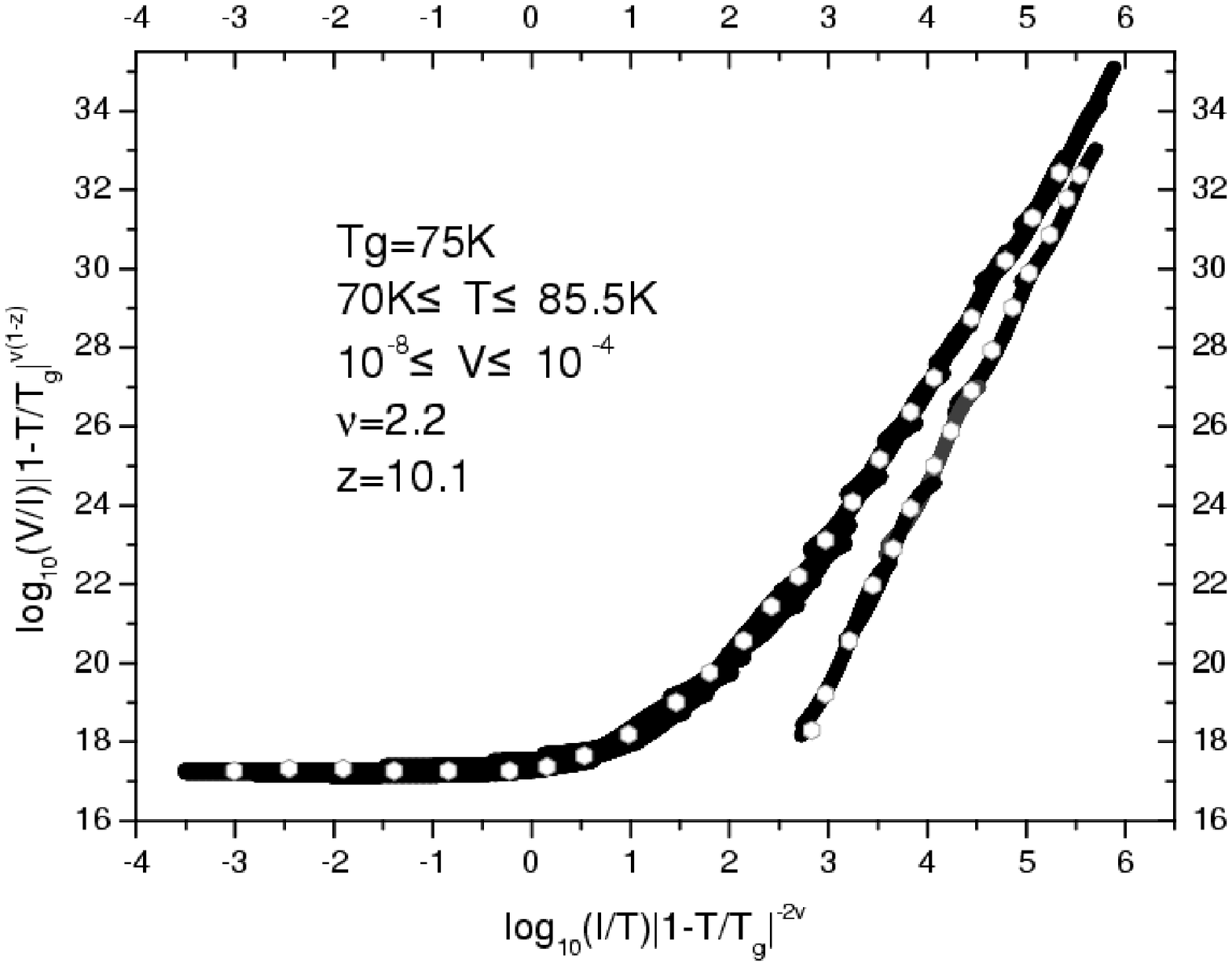}

\noindent (b)

\includegraphics[width=0.55\linewidth,angle=270]{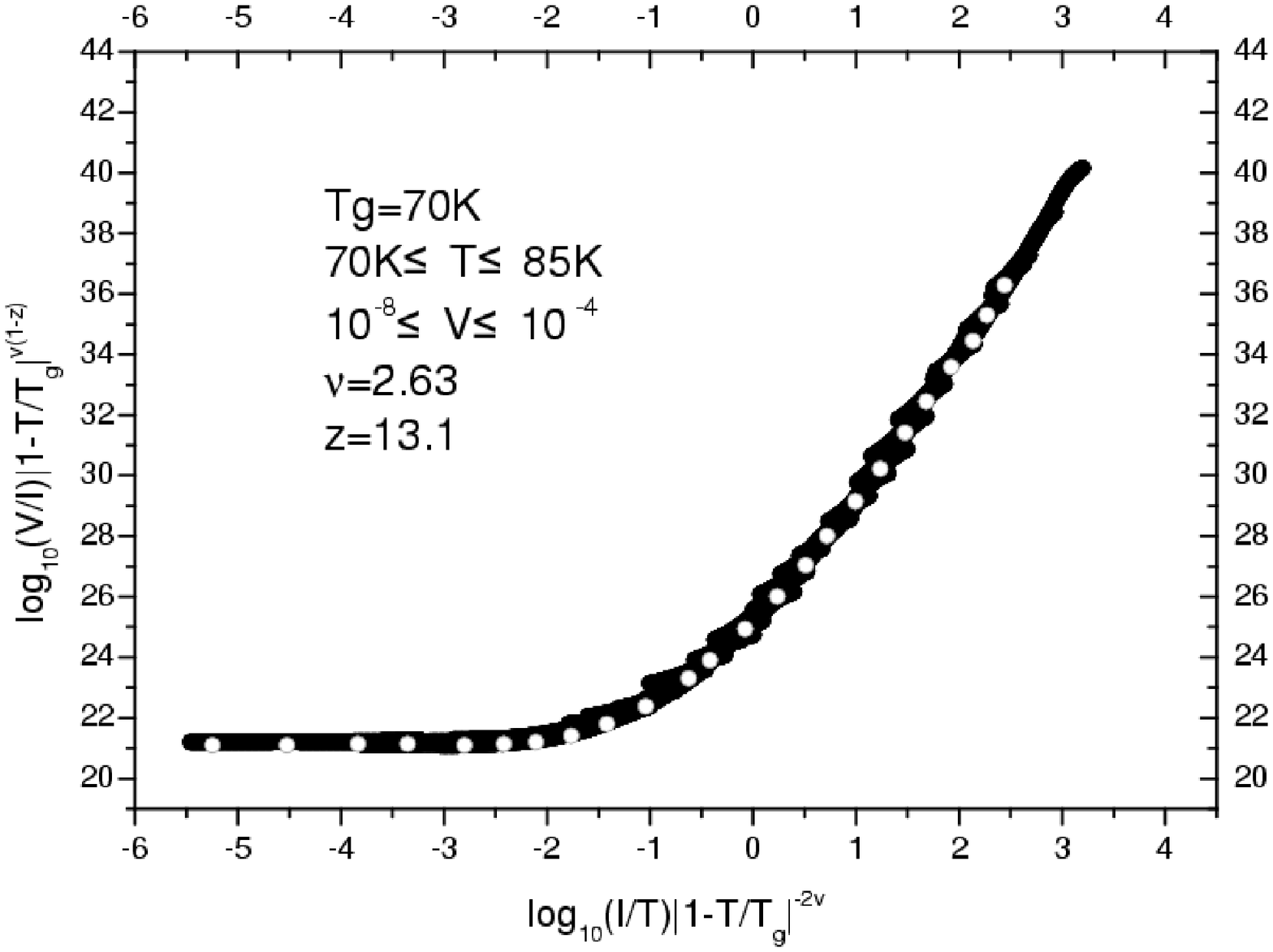}

\noindent (c)

\figcaption{$\bullet$: Collapse of the simulated I-V isotherms
in Fig.2(a) using various critical parameters compared with\\
$\circ$: the experimental I-V isotherms for a 2200$\text{\AA}$
YBCO film in 4T measured by Strachan et al.[3]. The conventional
analysis is shown in (a).\\}
\end{figure}
respectively. In Fig.3 we show the comparison of our result with
the data collapses of measured $I-V$ curves of Strachan et al. in
Ref.[3] and find fair agreement.

In Fig.3 of the Ref.[3] another serious puzzle is mentioned with
plotting the derivatives of the $ln{V}$ vs $ln{I}$ for
experimental isotherms following Repaci et al.\cite{R12}. All
isotherms seem to have a maximum slope at about some current with
Ohmic tails developing to the left of the peaks. This behavior is
inconsistent with the scaling function of vortex-glass
transition\cite{R4} which should predict a horizontal line in the
$[dln{V}/dln{I}]_T\thicksim ln{I}$ plot at $T=T_g$. Nevertheless,
our extended power law $I-V$ equations may give it a reasonable
explanation.

In Fig.4 we plot the slope of experimental $I-V$ isotherms of
Ref.[3] and [12] in an empirical scaling form
\begin{equation}\label{16}
\frac{S-1}{n-1}=\Phi(I/I_i)
\end{equation}
and find that these experimental data are collapsed, where the
subscript $i$ denotes the value at inflection point of $ln{V}\sim
ln{I}$ curve, $S\equiv[dln{V}/dln{I}]_{T}$ and $n$ denotes
$S_{max}$ at inflection. This empirical scaling function can be
derived from extended power law nonlinear I-V isotherms.
\begin{figure}
\centering \leftskip-5mm
\includegraphics[width=0.7\linewidth, angle=270]{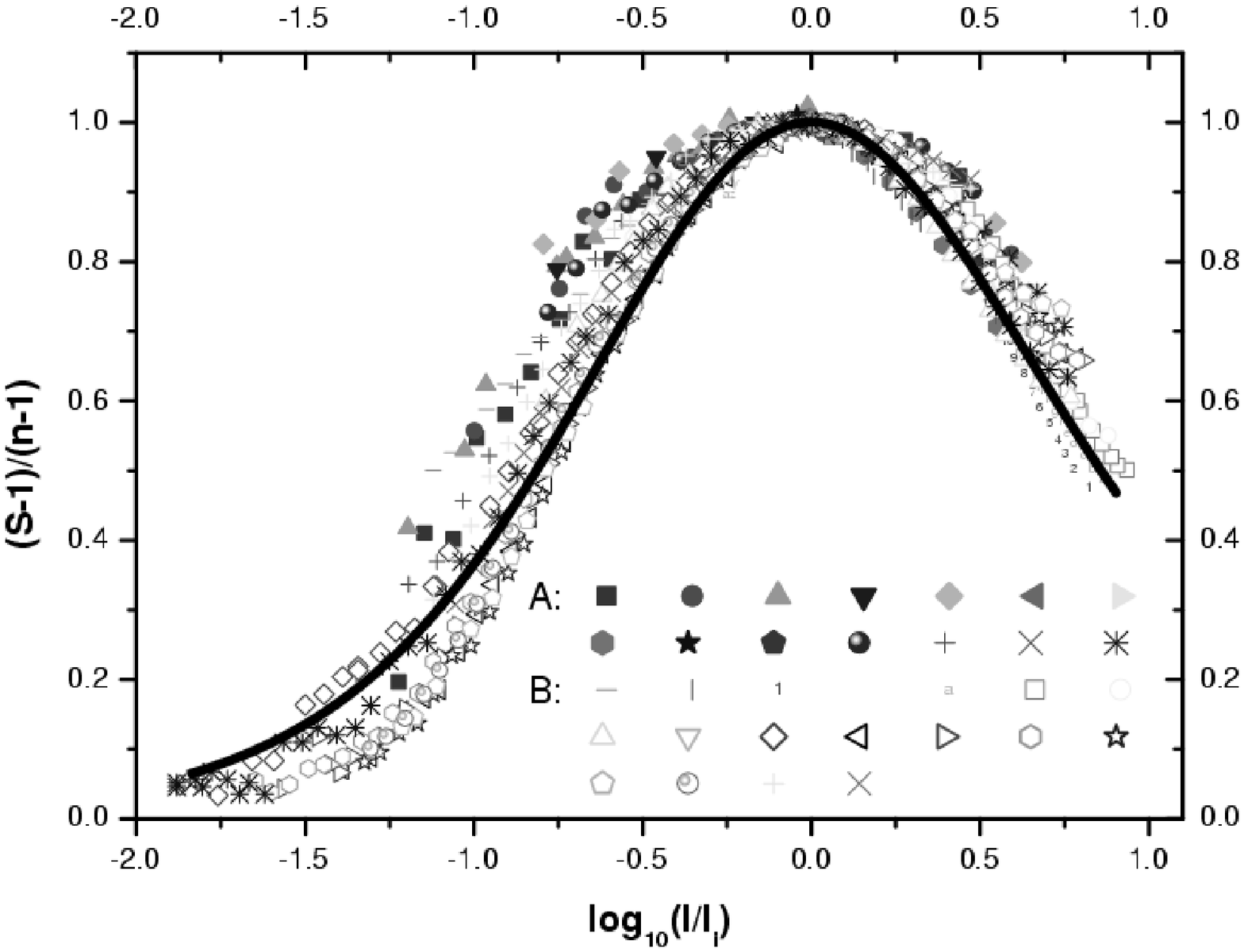}
\caption{Empirical scaling of the derivatives of the experimental
$lnV\sim lnI$ isotherms. A:data from Ref.12 B: data from Ref.3.
The solid curve is the results simulated from Eq.(17).}
\leftskip-5mm
\end{figure}
Starting from our equation (8) one finds analytic equation
\begin{equation}\label{17}
\frac{S-1}{n-1}=C\frac{(x-y)\ln{(x/y)}}{1+y-x+yp\ln{(x/y)}}
\end{equation}
where
\begin{equation}\label{17}
C=\frac{1+y_i-x_i+y_{i}p\ln{(x_i/y_i)}}{(x_i-y_i)\ln{(x_i/y_i)}}
\end{equation}
shown by the solid curve in Fig.4.

The agreement of the simulated I-V isotherms derived from Eqs.(8)
and (15) with the wide-range experimental data of Ref.[3] as shown
in our Fig.2,3 and 4 demonstrates that extended power law is the
general form of I-V isotherms. Though, in principle, each I-V
curve can have both positive and negative curvature parts joined
at the inflection point, the actually used current-voltage window
in measurements limits the observable part of each I-V isotherms
and results in the previous reasonable scaling analysis
\cite{R1,R2}. Using wider I-V windows of measurements, Strachan et
al. find that non-Ohmic-power-law-like behavior can be fit to all
$T<81K$ isotherms over at least four decades of voltage data and
thus insightfully reveal that the conventional I-V scaling cannot
determine a unique Tg with corresponding critical exponents. The
successful reproduce of their experimental data collapses with our
simulated I-V isotherms from equations (8) and (15) with the
uniquely assumed parameters the irreversibility temperature
$T^\ast(B=4T)=88K$ and $\alpha=3.0$, $\delta=0.5$, $p=1.75$,
$\gamma_0=3.21\times10^{21}$ shows that extended power law is
helpful in settling the controversy surrounding critical phenomena
in high temperature superconductors.

In conclusion, we show that dc current-voltage characteristic of
mixed state superconductors has the general form of extended power
law. Isotherms simulated from this nonlinear equation fit the
recent experimental I-V data of Strachan et al. This equation also
gives a reasonable explanation to the critical scaling relation
recently observed in the derivatives of the $lnV \text{ vs } lnI$.

This work is supported by the Ministry of Science \& Technology of
China (NKBRSG-G 1999064602) and the National Natural Science
Foundation of China under Grant No.10174003 and No.50377040.

\end{document}